\documentclass[a4paper,
               biblatex,     
               keeplastbox,   
               ]{jacow}
\usepackage{pdfpages,multirow,ragged2e} %
\makeatletter%
	\ifboolexpr{bool{xetex}}
	 {\renewcommand{\Gin@extensions}{.pdf,%
	                    .png,.jpg,.bmp,.pict,.tif,.psd,.mac,.sga,.tga,.gif,%
	                    .eps,.ps,%
	                    }}{}
\makeatother

\ifboolexpr{bool{xetex} or bool{luatex}} 
 {}                                      
 {\usepackage[utf8]{inputenc}}           

\usepackage[USenglish]{babel}

\newcommand{\q}[2]{\ensuremath{#1\ \mathrm{#2}}}

\begin{document}

\title{Symplectic Particle Tracking in a Thick Nonlinear \NoCaseChange{Mc}Millan Lens for the  Fermilab Integrable Optics Test Accelerator (IOTA)}
  

\author{B.~Cathey\thanks{bcathey@fnal.gov}, G.~Stancari, T.~Zolkin, Fermi
  National Accelerator Laboratory, Batavia, IL, USA}
	
\maketitle

\begin{abstract}
The McMillan system is a novel method to increase the tune spread of a beam without decreasing its dynamic aperture due to the system's integrability. While the ideal system is based on an infinitely thin kick, the physical design requires a thick electron lens, including a solenoid. Particle transport through the lens is difficult to simulate due to the nature of the force on the circulating beam. This paper demonstrates accurate simulation of a thick McMillan lens in a solenoid using symplectic integrators derived from Yoshida's method.
\end{abstract}

\section{INTRODUCTION}

As circular accelerators achieve higher beam intensities, beam instabilities increasingly limit design capabilities. Beam instabilities can come from interactions of the beam with the surrounding environment through wakefields and impedance, or from space charge due to the beam itself. Landau damping is one way to mitigate instabilities. Landau damping is the use of a spread of betatron tunes to lower a beam's sensitivity to instabilities. To generate a tune spread nonlinear forces are required, such as octupole magnets creating tune spread dependent on the particle’s amplitude. However, octupoles and other nonlinear elements can have a significant drawback in that they reduce the beam's dynamic aperture. There are nonlinear dynamical systems that are integrable, and that can be implemented in accelerators, without loss of dynamic aperture~\cite{Ruggiero:PA:1982, Danilov:SL-NOTE-94-74-AP:1994, Danilov:PAC:1997, Danilov:FN-671:1998, Danilov:JINST:2021, Danilov:PRSTAB:2010}.

The Integrable Optics Test Accelerator (IOTA) at Fermilab is partly dedicated to the experimental study of novel, integrable, nonlinear focusing lattices~\cite{Antipov:JINST:2017b}. In particular, one straight section is designed to include an electron lens, which will be used for research on nonlinear dynamics, electron cooling, and space-charge compensation~\cite{Stancari:AIP:2016, Antipov:JINST:2017b, Stancari:JINST:2021, Banerjee:cool2021}. Because of their flexibility, electron lenses can be designed to have different effects on the circulating beam~\cite{Shiltsev:PRL:2007, Stancari:PRL:2011, Fischer:PRL:2015, Shiltsev:elens-book:2016, Redaelli:JINST:2021}. In this paper, we focus on the simulation of a realistic McMillan electron lens, the nonlinear element necessary for creating the McMillan integrable system in IOTA \cite{McMillan:1971, Cathey:JINST:2021, cathey2021progress}.

\section{BACKGROUND}

The McMillan system has two sections: a linear transport and a nonlinear kick~\cite{Danilov:SL-NOTE-94-74-AP:1994, Danilov:PAC:1997}. The linear transport requires a $\pi/2$ phase advance in both vertical and horizontal phase spaces, and a round beam at the kick. The two dimensional McMillan kick is defined as
\begin{equation}
f(r) = \frac{k r}{\frac{r^2}{a^2} + 1},
\label{eq:kick}
\end{equation}
with $k$ as the kick strength and $a$ as an effective width. This system has two integrals of motion guaranteeing integrability. These are best defined in polar coordinates where:
\begin{equation}
\begin{aligned}
r &= \sqrt{x^2 + y^2} , & r' &= \frac{x x' + y y'}{\sqrt{x^2 + y^2}} , \\
\theta &= \arctan2{(y,x)} , & \theta' &= \frac{x y'
  - y x'}{\sqrt{x^2 + y^2}} .
\end{aligned}
\label{eq:polar}
\end{equation}
One is the angular momentum $L_z = r \theta'$. The other is given by
\begin{equation}
  I = \frac{r^2 r'^2}{a^2} +
    \frac{r^2}{\beta^2} + r'^2 + \theta'^2 - k r r' .
\label{eq:I-const}
\end{equation} 
This system generates a wide tune spread while maintaining integrability~\cite{Cathey:JINST:2021, Nagaitsev-Zolkin:PRAB:2020}. There are two regimes of strength for the McMillan lens based on the lattice beta function. The weak regime occurs when $\beta k < 2$ and has a maximum tune spread of
\begin{equation}
  (\Delta \nu_\theta)_{\mathrm{max}} = \frac{1}{4} - \frac{1}{2\pi}
  \arccos{\left(\frac{\beta k}{2}\right)}.
  \label{eq:max-detuning-weak-lens}
\end{equation}
The strong regime is when $\beta k > 2$ and can create a tune spread that reaches the integer resonance.

\begin{figure}
  \centering
  \includegraphics[width=\columnwidth]{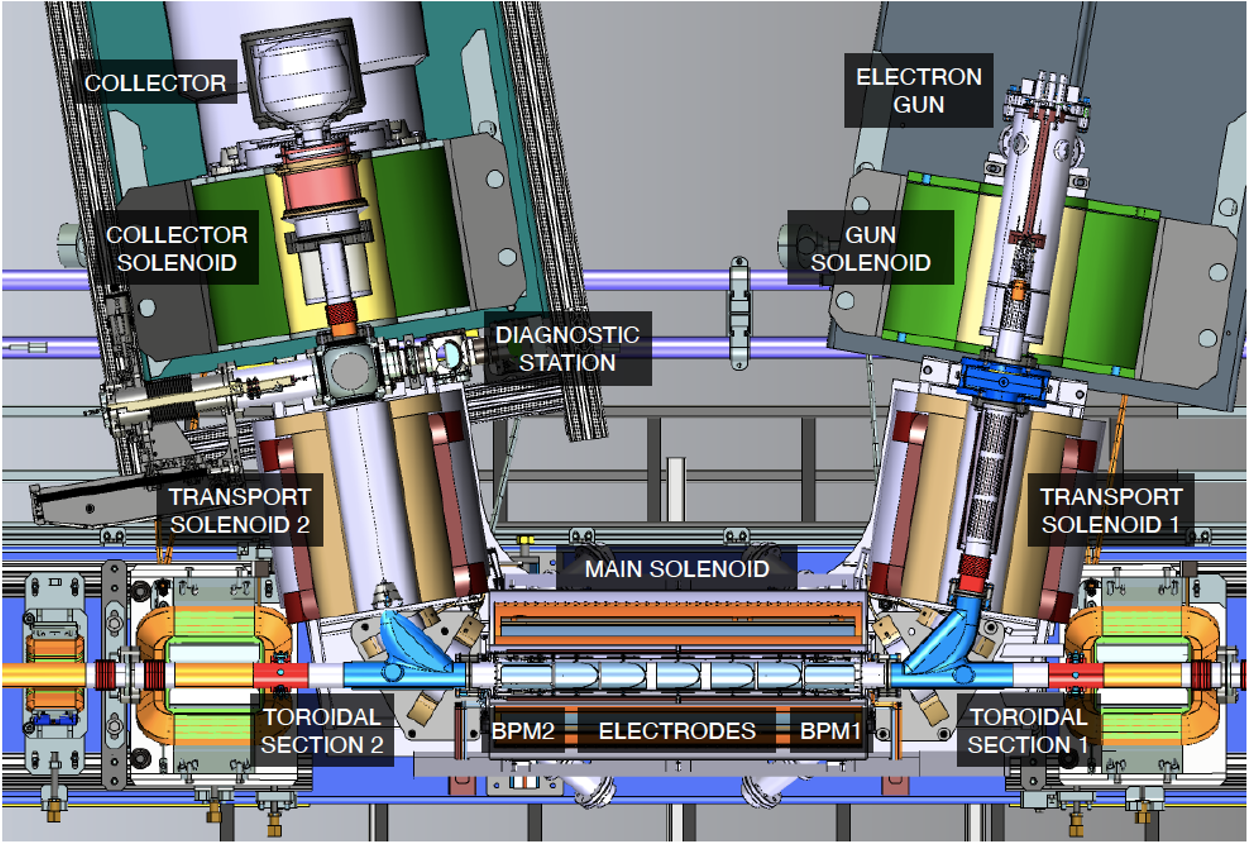}
  \caption{Drawing of the electron lens in IOTA.}
  \label{fig:MOPA17f1}
\end{figure}

However, this system cannot be physically achieved with a thin kick. The electron lens must have some length and be contained within a solenoid to magnetically confine the electron beam and keep the lens current consistent throughout beam traversal. These complications are shown in Fig.~\ref{fig:MOPA17f1}. To predict how they affect particle motion compared to the ideal system, the motion through the lens needs to be simulated.

\section{SYMPLECTIC INTEGRATION OF THE MCMILLAN LENS}

The most straightforward way to simulate the lens is to solve the equations of motion for a particle in the lens. To do so, the lens is analyzed in isolation from the rest of the lattice, treating the lens beam as a current density with infinite length. The kick in Eq.~\eqref{eq:kick} is created with a current density of the form
\begin{equation}
  J(r) = \frac{j_0}{\left( \frac{r^2}{a^2} + 1 \right)^2} ,
\label{eq:density}
\end{equation}
where $j_0$ is the peak current density on axis. The current density gives a total force from the electric and magnetic fields of
\begin{equation}
  \vec{F}(r) = \frac{e \left(1 \pm \beta_z \beta_e \right) j_0}{2 \epsilon_0 \beta_e c} \frac{r}{\frac{r^2}{a^2} + 1} \hat{r} = \kappa \frac{r}{\frac{r^2}{a^2} + 1} \hat{r}  ,
\label{eq:force}
\end{equation}
with $\beta_e$ and $\beta_z$ being the ratios to $c$ of the lens beam's velocity and the circulating beam's velocity respectively. This force has a potential energy of the form
\begin{equation}
  V = \frac{-\kappa a^2}{2} \ln \left( \frac{r^2}{a^2} + 1 \right).
\label{eq:potent}
\end{equation}
The solenoid is treated as a uniform magnetic field, $B$, along the beam axis. The fringe effects of the solenoid can then be added before and after the main lens transport. This isolated lens system has two constants of motion as well: the angular momentum and the total energy given by
\begin{equation}
\begin{aligned}
L_M &= p_z r \theta' - \frac{e B r^2}{2} ,\\
E_M &= \frac{p_z^2}{2 m} \left(r'^2 + \theta'^2 \right) - \frac{\kappa a^2}{2} \ln \left( \frac{r^2}{a^2} + 1 \right).
\end{aligned}
\label{eq:lens_consts}
\end{equation}
Here, $p_z$ is the momentum of the circulating beam particle along the beam axis, and $m$ is its mass.

Unfortunately, logarithmic central potentials such as Eq.~\eqref{eq:potent} have no known solutions. As such, the thick lens must be approximated. While the final phase space position cannot be directly calculated, we can see how close to the correct value the approximation is based on the error to the constants in Eq.~\eqref{eq:lens_consts} after transport through the lens. The simplest approach is to use alternating solenoid transports and kicks of the form of Eq.~\eqref{eq:kick}. However, increasing numbers of transport-kick slices approached convergence very slowly. A super convergent method is needed to accurately and efficiently simulate the McMillan lens.

As the lens system has two separable Hamiltonians, the lens beam and the solenoid, Yoshida's method for constructing symplectic integrators is viable \cite{Yoshida_1990}. Yoshida's method creates coefficients for increasing orders of integrator. These coefficients are applied to the length, $L$, of the transport or kick. The kick coefficient, $k$, is given by
\begin{equation}
  k = \frac{L}{p_z \beta_z c} \kappa = \frac{L \left(1 \pm \beta_z \beta_e \right) j_0}{2 (B \rho) \epsilon_0 \beta_z \beta_e c^2},
\label{eq:k}
\end{equation}
so that the kick strength is simply multiplied by the appropriate coefficient. The solenoid transport matrix has its effective length multiplied by the coefficient \cite{chao2013handbook}. The second order integrator used as a base for the Yoshida method was: transporting half-way through the solenoid, a full strength kick, and transporting another solenoid half $\left( \frac{1}{2}, 1, \frac{1}{2} \right)$.

\begin{figure}
  \centering
  \includegraphics[width=\columnwidth]{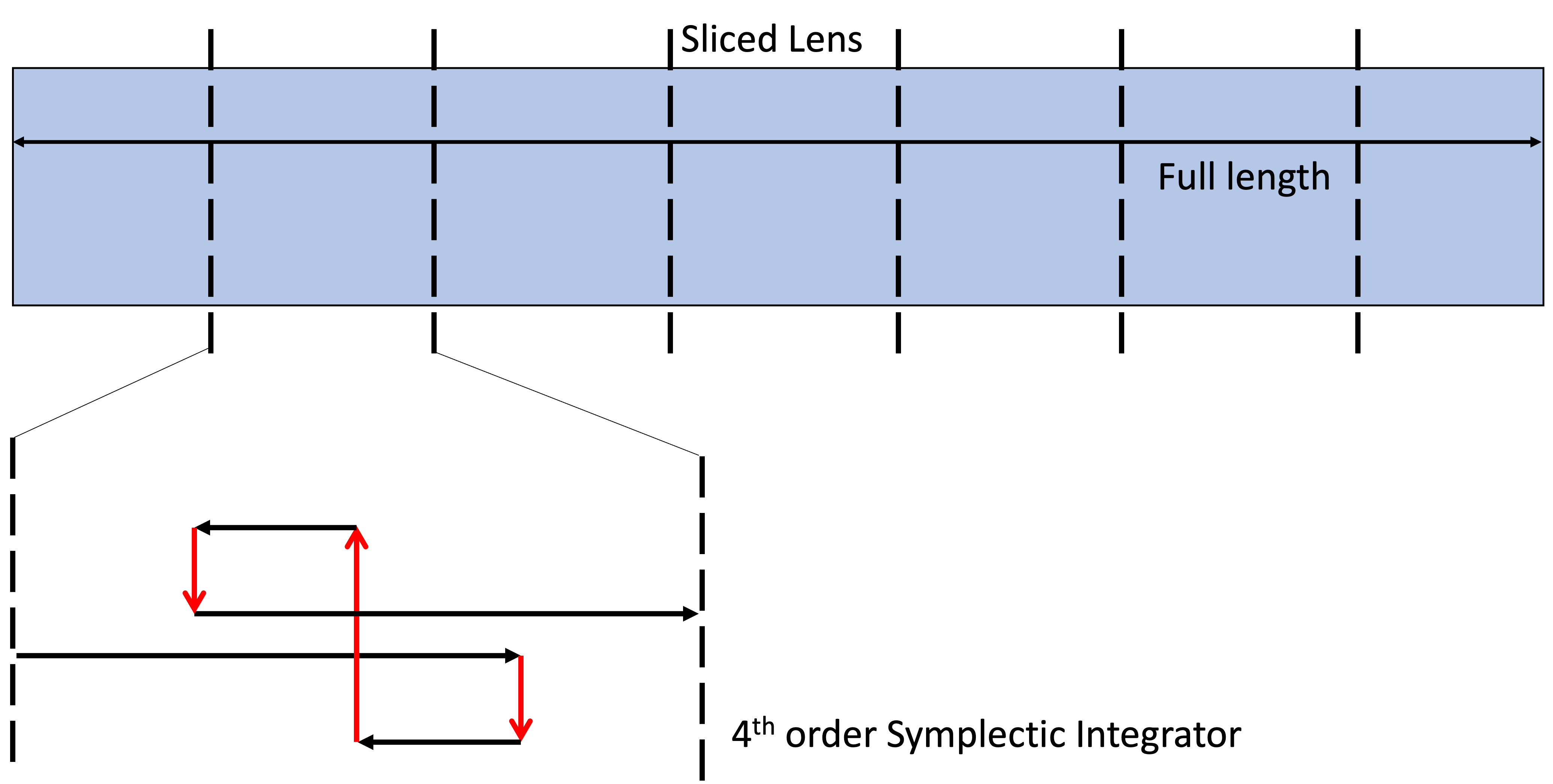}
  \caption{Diagram showing how the lens is approximated using a symplectic integrator over equal slices of the lens.}
  \label{fig:MOPA17f2}
\end{figure}

To further increase accuracy, the lens was first sliced into small equal units of length, and each individual unit was symplectically integrated over its smaller length. Figure \ref{fig:MOPA17f2} shows an example with the fourth order integrator. This was done for multiple orders of integrator.

\section{SIMULATIONS OF THE MCMILLAN LENS}

\begin{figure}
  \centering
  \includegraphics[width=\columnwidth]{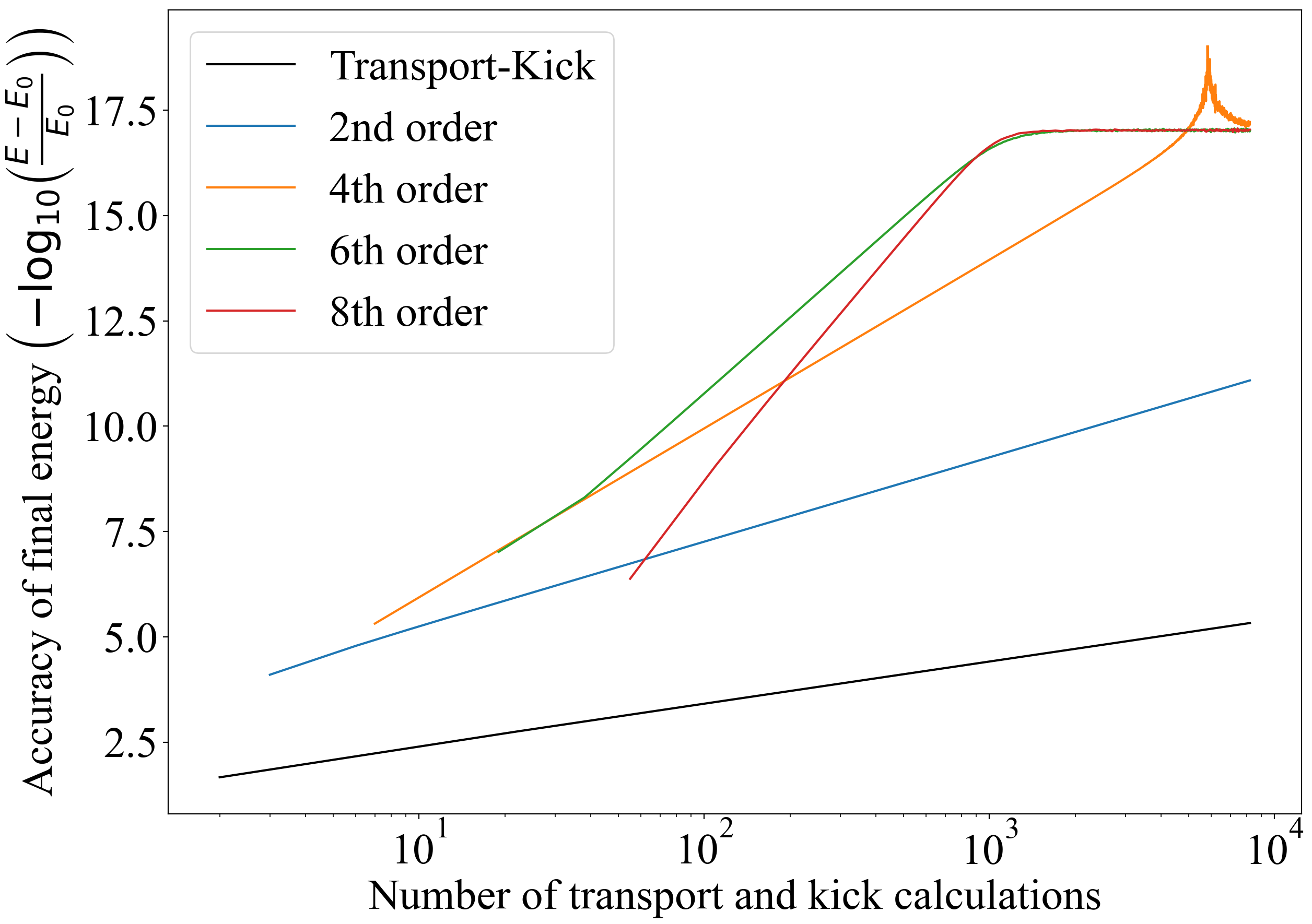}
  \caption{Graph showing how different orders of integrator converge with increasing numbers of slices. The particle started with $x, x', y, y' = (\q{2}{mm}, \q{-0.1}{mrad}, \q{-3}{mm}, \q{0.4}{mrad})$.}
  \label{fig:MOPA17f3}
\end{figure}

Simulations were conducted using an isolated lens with a length of $L = \q{0.7}{m}$, kick strength $k = \q{1.076}{ m^{-1}}$, and solenoid field $B = \q{0.333}{T}$. Figure~\ref{fig:MOPA17f3} shows an example of how different orders converge on the energy for a test particle. Only the energy convergence is shown because $L_M$ quickly converged. The results show how the symplectic integrators converge much faster than the transport-kick approximation, eventually reaching machine accuracy.

\begin{figure}
  \centering
  \includegraphics[width=\columnwidth]{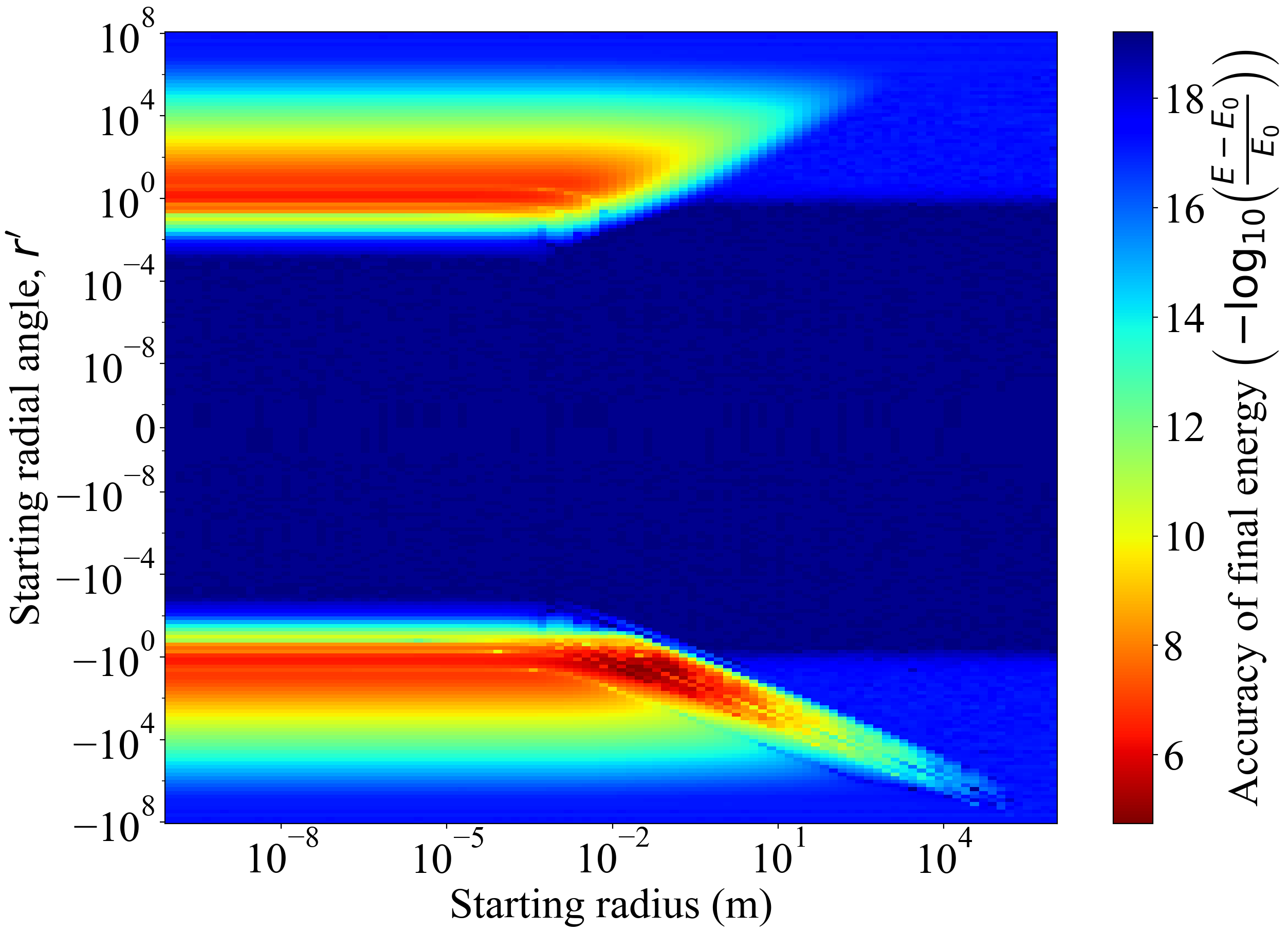}
  \caption{Color plot showing the error in energy for 50 slices with the 6th order integrator for the $\theta'=0$ phase space.}
  \label{fig:MOPA17f5}
\end{figure}

The rate of convergence for each integrator depends on the initial phase-space coordinates of a particle. For simplicity's sake, a single approximation that adequately covers all relevant particle starting coordinates needs to picked. After looking at a few sample starting positions, the sixth order integrator looked like a good choice for accurate results. Figure~\ref{fig:MOPA17f5} shows the error in total energy after transporting through the lens for 50 slices of the sixth order with a phase space where the initial $\theta' = 0$. A very wide range of starting radii and radial velocities was chosen to explore the behavior of this integration scheme. The plot shows that accuracy decreases in regions with specific radial velocities. These regions are at the edge of typical beam parameters and will only affect larger beam sizes. Particles outside of these regions were transported with high accuracy.

\begin{figure}
  \centering
  \includegraphics[width=\columnwidth]{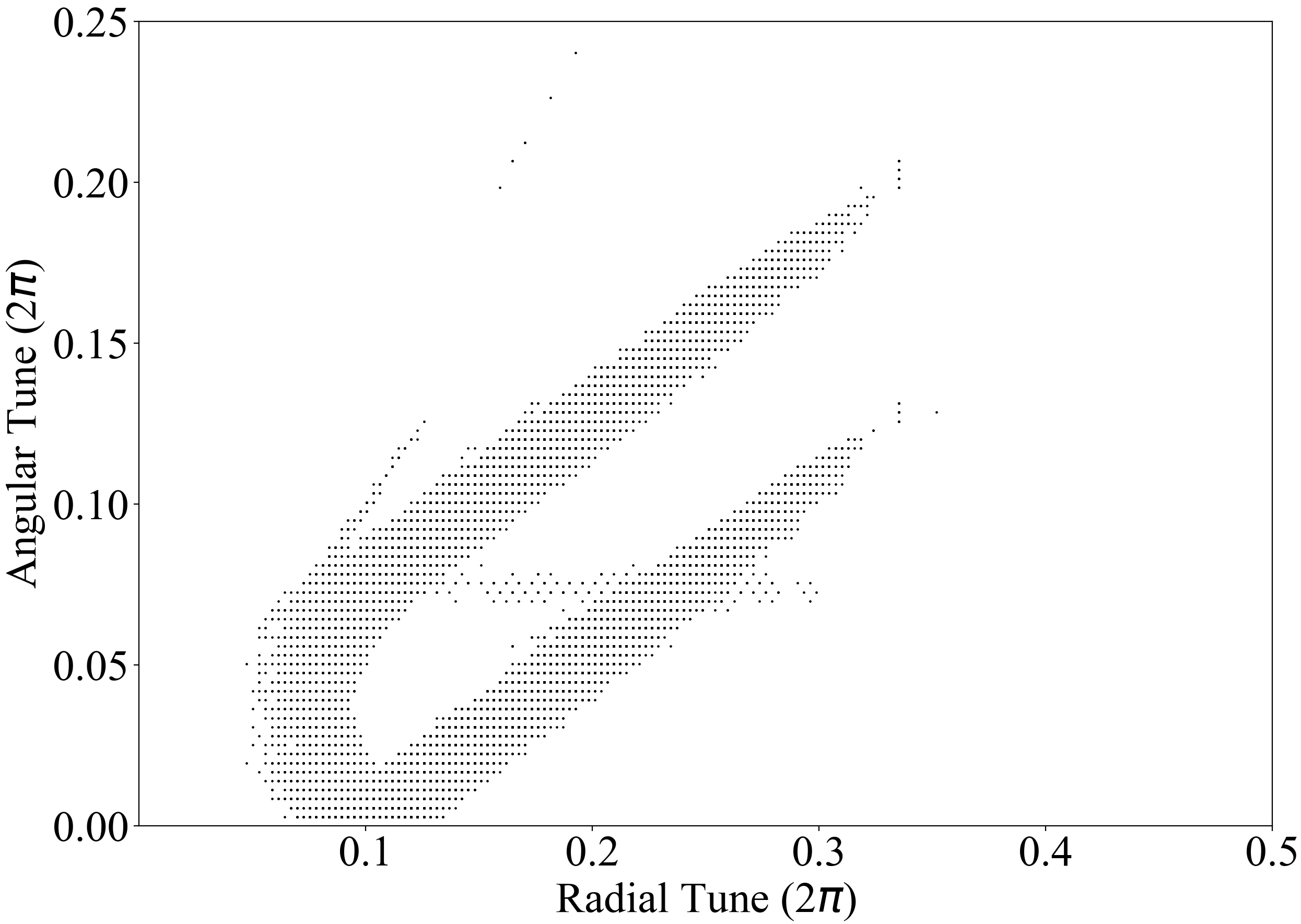}
  \caption{Plot of the radial and angular tune spread of a weak ($k=\q{0.538}{m^{-1}}$) lens. The beam was tracked for 360 turns.}
  \label{fig:MOPA17f6}
\end{figure}

\begin{figure}
  \centering
  \includegraphics[width=\columnwidth]{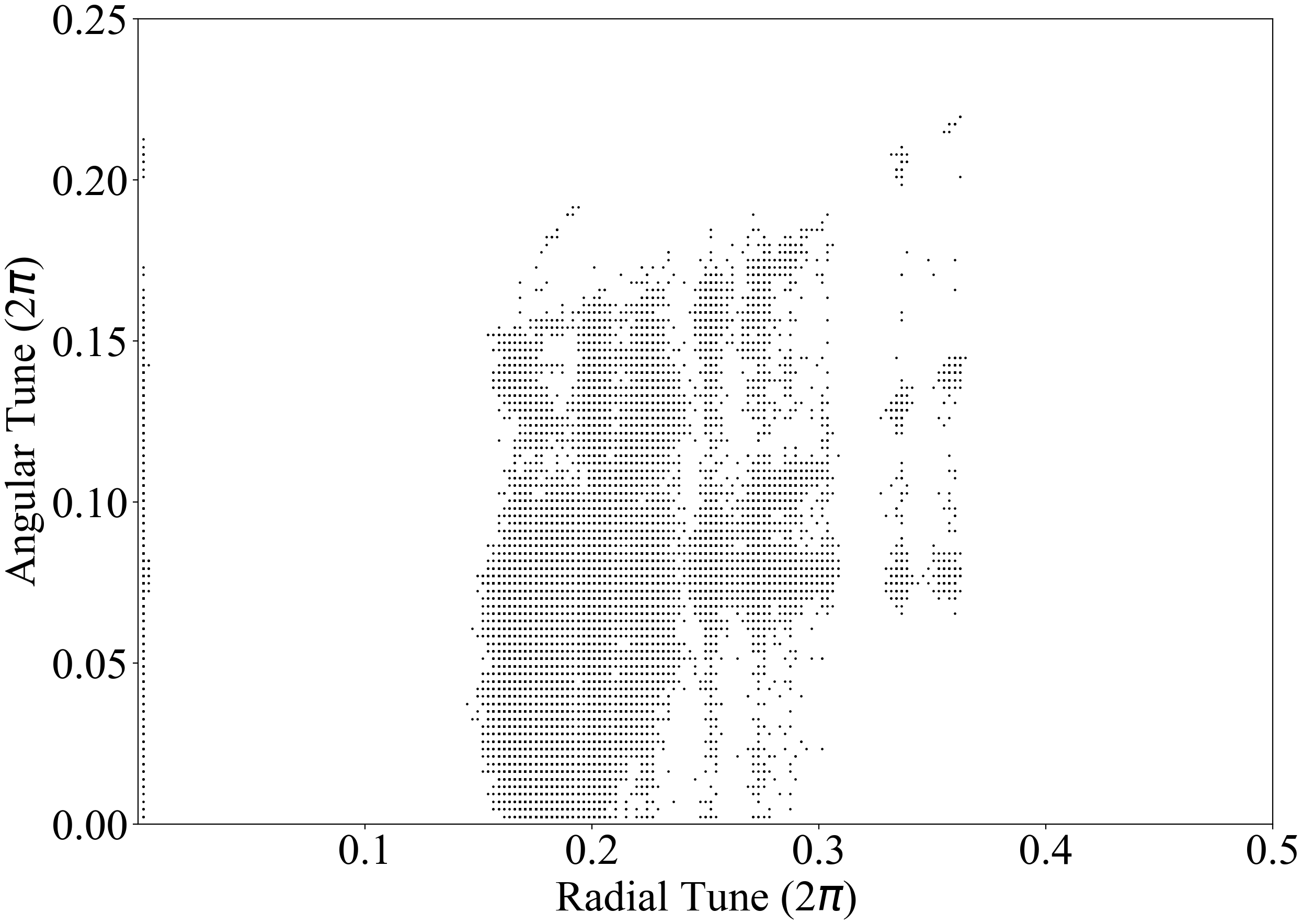}
  \caption{Plot of the radial and angular tune spread of a strong ($k=\q{1.076}{m^{-1}}$) lens. The beam was tracked for 430 turns.}
  \label{fig:MOPA17f7}
\end{figure}

Using the above approximation, initial studies of an orbiting beam in the full ring were done using a simple transport matrix with a $\pi/2$ phase advance, as required by the ideal McMillan system, with $\beta = \q{3}{m}$ at the lens entrance. The beam had an uncorrelated Gaussian distribution with initial horizontal and vertical emittances of 2.95~mm~mrad. Figures~\ref{fig:MOPA17f6} and~\ref{fig:MOPA17f7} show the tune spreads due to a weak and strong lens, respectively. Interestingly, the weak lens reached the integer angular resonance, which is not possible in the ideal McMillan system.

\section{CONCLUSIONS}

This method allows for more realistic simulations of the McMillan electron lens. As the approximation can be computationally expensive, running simulations on parallel processors will be necessary to look at long term motion of the beam. As such, we are planning to implement this method in a tool with parallel processing. Further additions can be done to improve modeling of the system. One possibility is adding space-charge both outside and inside the lens to see how stability and tune shifts are affected, and another is electron lens imperfections to see how sensitive the system is. Finally, there is the question of how to best design and implement experiments in IOTA. As the system now deviates from integrability, is there a way to adjust IOTA such that this deviation can be minimized? Removing the phase advance due to lens focusing from the $\pi/2$ phase advance outside the lens, or adding a compensating solenoid elsewhere might bring the beam closer to the ideal case. Now we can test these and other cases to determine the best way to proceed experimentally.

\section{ACKNOWLEDGMENTS}

This manuscript has been authored by Fermi Research Alliance, LLC under Contract No. DE-AC02-07CH11359 with the U.S. Department of Energy, Office of Science, Office of High Energy Physics.  Report no.~FERMILAB-CONF-22-549-AD.

\printbibliography

\end{document}